\documentclass[12pt]{iopart}
\usepackage{bm}
\usepackage{graphicx}
\usepackage{setstack}
\usepackage{amsthm}
\usepackage{amssymb}
\usepackage{amsfonts}
\usepackage{ifthen}
\usepackage{cite}
\renewcommand{\vec}{\bm}
\newcommand{\js}{\vec{j}_{\mathrm{s}}} 
\newcommand{\Lfp}{\ensuremath{\mathcal{L}_{0}}} 
\newcommand{\kb}{k_{\mathrm{B}}} 
\newcommand{\Lone}{\ensuremath{\mathcal{L}_{1}}} 
\newcommand{\sgn}{\ensuremath{\mathop{\mathrm{sgn}}}}
\newcommand{\tstot}{\ensuremath{\Delta \tilde{s}_{\mathrm{tot}}}}  
\newcommand{\stot}{\ensuremath{\Delta s_{\mathrm{tot}}}} 
\newcommand{\nt}{\ensuremath{\tilde{n}}}
\newcommand{\nb}{\ensuremath{\hat{n}}}
\newcommand{\wt}{\ensuremath{\tilde{w}}}
\newcommand{\ps}{\ensuremath{p_{\mathrm{s}}}}
\newcommand{\nut}{\ensuremath{\tilde{\vec{\nu}}}}
\newcommand{\muv}{\ensuremath{\vec{\mu}}}
\newcommand{\nuv}{\ensuremath{\vec{\nu}}}

\newcommand{\dtp}{\ensuremath{\mathop{\mathrm{d}t'}}}

\newcommand{\lt}{\ensuremath{\lambda_{0}}}
\newcommand{\Lc}{\ensuremath{\mathcal{L}}}
\newcommand{\xt}{\ensuremath{\tilde{\vec{x}}}}
\newcommand{\xv}{\ensuremath{\vec{x}}}
\newcommand{\xh}{\ensuremath{\hat{\vec{x}}}}
\newcommand{\At}{\ensuremath{\tilde{\mathcal{A}}}}
\newcommand{\Ac}{\ensuremath{\mathcal{A}}}

\newcommand{\Gp}{\ensuremath{\Gamma_{\pi}^{+}}}  
\newcommand{\Gm}{\ensuremath{\Gamma_{\pi}^{-}}}  
\newcommand{\nf}{\ensuremath{\mathfrak{n}_{\pi} }}
\newcommand{\mf}{\ensuremath{\mathfrak{m}_{\pi} }}

\newcommand{\lax}{\ensuremath{\lambda_{x}}} 
\newcommand{\Cc}{\ensuremath{\mathcal{C}}}
\newcommand{\nv}{\ensuremath{\vec{n}}}
\begin{document}
\title[Apparent Entropy Production in Networks with Hidden Slow Degrees of Freedom]
{Fluctuations of Apparent Entropy Production in Networks with Hidden Slow Degrees of Freedom}
	\author{Matthias Uhl, Patrick Pietzonka and Udo Seifert}
	\address{II. Institut f\"ur Theoretische Physik, Universit\"at Stuttgart, 70550
	Stuttgart, Germany}

	\begin{abstract}
		The fluctuation theorem for entropy production is a remarkable symmetry
		of the distribution of produced entropy that holds universally in
		non-equilibrium steady states with Markovian dynamics.  However, in
		systems with slow degrees of freedom that are hidden from the observer,
		it is not possible to infer the amount of produced entropy exactly.
		Previous work suggested that a relation similar to the fluctuation
		theorem may hold at least approximately for such systems if one
		considers an apparent entropy production. By extending the notion of
		apparent entropy production to discrete bipartite systems, we
		investigate which criteria have to be met for such a modified
		fluctuation theorem to hold in the large deviation limit.  We use
		asymptotic approximations of the large deviation function to show that
		the probabilities of extreme events of apparent entropy production
		always obey a modified fluctuation theorem and, moreover, that it is
		possible to infer otherwise hidden properties. For the paradigmatic
		case of two coupled colloidal particles on rings the rate function of
		the apparent entropy production  is calculated to illustrate this
		asymptotic behavior and to show that the modified fluctuation theorem
		observed experimentally for short observation times does not persist in
		the long time limit.
	\end{abstract}
	\newpage
	\section{Introduction}
	Stochastic thermodynamics has proven to be a reliable framework for the
	description of systems far from equilibrium. By extending the notion of
	classical thermodynamics to the level of single trajectories in a well
	defined way it allows us to identify laws from macroscopic thermodynamics
	like conservation of energy or the second law on the mesoscopic
	scale~\cite{jarzynski_equalities_2011, seifert_stochastic_2012,
	vandenbroeck_ensemble_2015}.
	One important class of such relations are fluctuation theorems that
	constrain the probability distributions of fluctuating quantities like the work
	produced along a trajectory.
	A prominent example is the fluctuation theorem (FT) for the total entropy
	production in a non equilibrium steady state (NESS)
	\begin{equation}
		\ln \frac{p(\stot)}{p(-\stot)} = a \stot 
		\label{eq:fluc_theorem}
	\end{equation}
	with $a=1$, and we set $\kb=1$ throughout this paper. It states that
	the probability of negative entropy production is exponentially suppressed and
	implies a positive average entropy production. It can
	therefore be seen as a refinement of the second law of thermodynamics to
	mesoscopic scales where fluctuations are still relevant.

	For Markovian dynamics on a finite state space or finite dimensional driven
	Brownian diffusion, the FT follows as a mathematical identity, relying only
	on some rather technical conditions like existence and uniqueness of the
	stationary state. In addition, for a physical identification of the total
	entropy production, it is important that one has identified and taken into
	account all relevant slow degrees of freedom such that there is a clear
	time-scale separation between them and all other degrees of freedom. In
	this case the system evolves according to a Markovian dynamics described by
	a Langevin or Master equation.

	This assumption is certainly fulfilled for the paradigmatic example
	of a Brownian particle in an aqueous solution serving as a heat bath, since it
	is safe to assume that collisions of the colloid with the molecules of the
	solvent happen on a different time-scale than the motion of the particle
	itself.
	However, the distinction between slow and fast degrees of freedom is often
	not that obvious, which has given rise to systematic studies of stochastic
	thermodynamics under an effective, coarse grained
	dynamics~\cite{rahav_fluctuation_2007, gomez-marin_lower_2008,
	puglisi_entropy_2010, esposito_stochastic_2012,
	altaner_fluctuation-preserving_2012, bo_entropy_2014, knoch_cycle_2015,
	zimmermann_effective_2015, frenzel_coarse-grained_2016}.
	A fundamental problem in this context is the
	description of systems for which individual states are not discernible, so
	that an external observer can obtain only partial information about the
	trajectory in state space~\cite{roldan_estimating_2010,
	polettini_effective_2017, shiraishi_fluctuation_2015}.
	An understanding of these questions can lead to inference methods that
	allow the observer to reconstruct the hidden properties of the
	system~\cite{flomenbom_utilizing_2006, amann_communications_2010,
	roldan_estimating_2010, alemany_free_2015, bravi_inferring_2017}.
	A recent study~\cite{strasberg_stochastic_2017} emphasizes the conceptual
	similarities between a system with slow hidden degrees of freedom and a
	system coupled strongly to a heat bath~\cite{seifert_first_2016,
	jarzynski_stochastic_2017}. 
	
	In the present paper we focus on systems with several slow degrees of
	freedom.  The state space of such a system is the product space of the
	states of the individual degrees of freedom. Hence, if one of the degrees
	of freedom is deemed hidden from the
	observer~\cite{crisanti_nonequilibrium_2012, gupta_fluctuation_2016,
	ehrich_stochastic_2017}, huge parts of this product space become
	indiscernible.
	We investigate the effects of such hidden slow degrees of freedom on the
	fluctuation theorem for entropy production in the stationary state.
	More precisely, we use large deviation theory to analyze which conditions
	have to be met for a modified FT~\eref{eq:fluc_theorem} with $a \neq 1$ to
	hold for the apparent entropy production obtained from an appropriate
	marginalization over slow hidden degrees of freedom
	in the limit of long observation times. 

\begin{figure}
		\centering
		\includegraphics[scale=1]{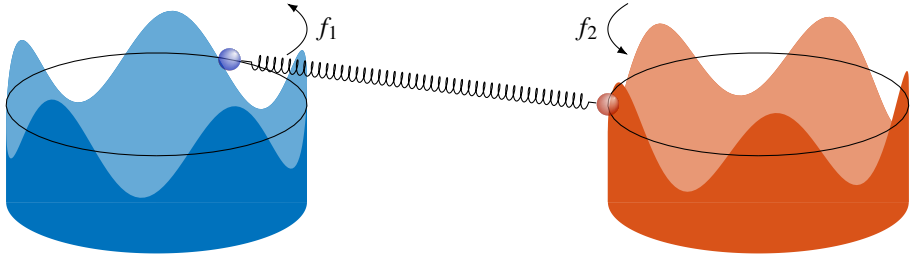}
		\caption{Two colloidal particles on rings coupled by an interaction
			force. If only one of the two is observable this setup serves as an
			example of a system with a slow hidden degree of freedom. Since the
			dynamics of the visible colloid is not Markovian the apparent
			entropy production of the visible part will in general not satisfy
			a FT. }
		\label{fig:colloids_on_rings}
	\end{figure}

	The experimental study performed by Mehl et~al.\cite{mehl_role_2012} on a
	system of two coupled colloidal particles on
	rings, which could also be seen as a toy model for synchronization, see
	e.g.~\cite{uchida_generic_2011, kotar_optimal_2013,
	izumida_energetics_2016}, has hinted at the approximate validity of the FT
	for a system consisting of two coupled colloidal particles on rings in
	aqueous solution that were each driven out of equilibrium by constant
	forces, as it is shown schematically in \fref{fig:colloids_on_rings}. While
	the combined system is still Markovian, the dynamics of an individual
	particle is not, since a time-scale separation is not valid for the two
	equally sized colloids.
	Naively inserting the visible dynamics into the definition of total entropy
	production in Markovian systems leads to an apparent entropy production,
	which was found to satisfy a modified FT \eref{eq:fluc_theorem} with $a
	\neq 1$ in a substantial range of parameters. However, this result can only
	be an approximation in some sense, since simulations of the system have
	revealed that there exist parameter combinations for which this relation
	does not hold and, hence, the right hand side of~\eref{eq:fluc_theorem} is
	replaced by some non-linear antisymmetric function. 

	The aim of the present paper is to shed light on this behavior and to
	identify the criteria that have to be met for such a modified FT to hold.
	Since  experimental data is by design limited to a finite number of
	experiments and large negative entropy production is rare, the modified FT
	has only been tested in a small region around zero
	in~\cite{mehl_role_2012}. Therefore, these results could be explained simply
	by the fact that every antisymmetric function is linear around the origin.
	To overcome these limitations and to check whether similar results also
	hold in the long time limit and for untypical realizations of the apparent entropy
	production, we use a large deviation approach to this problem.
	
	This paper is structured as follows. In~\sref{sec:apparent_entropy}, we
	revisit the definitions used in~\cite{mehl_role_2012} for systems with
	continuous degrees of freedom and define the apparent entropy production
	for a system with a discrete set of states. 
	In~\sref{sec:bipartite_grid}, we apply these definitions to a simple system
	with discrete states that is inspired by the one used in the experiments.
	We identify parameter values for which the rate functions of the apparent
	entropy production approximately obeys a modified FT even when the
	observable and hidden degree of freedom are strongly coupled.
	The explanation of this behavior is presented
	in~\sref{sec:bounds_and_approximations}, where we derive a bound on the
	rate function that becomes tight for large deviations from the mean and for
	which a modified FT holds. Based on an approximation to the characteristic
	polynomial of the so called tilted operator we show why in most cases the
	modified FT is not limited to the asymptotics of the rate function.
	In~\sref{sec:continuous_case}, we calculate the rate function of systems of
	the type used by Mehl et~al. in order to check whether our findings
	persist in the continuous limit. It turns out that the tails of the rate
	function still obey a modified FT exactly, while the
	deviations for smaller apparent entropy production rates are more
	pronounced than in the discrete case since the aforementioned approximation
	of the characteristic polynomial increasingly fails in the continuum limit case.
	Finally, we conclude in~\sref{sec:conclusion}.

	\section{Apparent Entropy Production}
	\label{sec:apparent_entropy}

	In systems with hidden degrees of freedom it is in general not possible to
	infer the total entropy production along a trajectory, since not all
	transitions are observable. In \cite{mehl_role_2012}, the notion of an apparent
	entropy production was introduced that can be calculated from the
	observable dynamics and which is equivalent to the total entropy production if
	all slow degrees of freedom are observable.
	
	In this section, we will shortly reproduce the definitions introduced in
	\cite{mehl_role_2012} for systems with continuous degrees of freedom and
	extend them to arbitrary discrete networks.

	\subsection{Continuous degrees of freedom}
	We assume that the time evolution of the probability density $p(\xv,t)$ to
	find the system in state $\xv$ at time $t$ is described by the
	Fokker-Planck equation
	\begin{equation}
		\partial_{t} p(\xv,t) = - \nabla \vec{j}(\xv,t)   
	\end{equation}
	with the probability current
	\begin{equation}
		\vec{j}(\xv,t) \equiv \mu \vec{F}(\xv) p(\xv,t) -D
		\nabla p(\xv,t)\,,
		\label{eq:prob_current}
	\end{equation}
	where $\vec{F}(\xv)$ is the force acting on the system, $D$ denotes the
	diffusivity, and $\mu = D/T$ is the mobility.
	From now on we use reduced units for time, space, energy, and temperature
	that allow us to set $D$, $\mu$, and $T$ to unity.
	The system will relax into a stationary state $\ps(\xv)$ for which the
	associated stationary current $\js(\xv)$ is divergence free.
	In a system with continuous degrees of freedom, the total entropy
	production up to time $t$ can be calculated by integrating the product of
	the local mean velocity $\nuv(\xv) \equiv \js(\xv)/\ps(\xv)$ with the
	fluctuating velocity along the trajectory,
	\begin{equation}
		\stot = \int_{0}^{t} \nuv(\xv(t')) \cdot \dot{\xv}(t') \dtp\,.
		\label{eq:stot_cont}
	\end{equation}

	Now we assume that not all degrees of freedom are visible, i.e.\,, the
	state $\xv$ is comprised of an observable part $\xt$ and a hidden part
	$\xh$. Thus it is no longer possible to calculate the full mean
	velocity from observation of the visible degrees of freedom. Instead, only
	the observable components of the mean velocity $\nut(\xt)$ conditioned on the
	visible state are accessible.

	The apparent entropy is therefore defined in analogy to~\eref{eq:stot_cont}
	as the integral
	\begin{equation}
		\tstot \equiv \int_{0}^{t} \nut(\xt(t')) \cdot \dot{\xt}(t') \dtp\,.
	\end{equation}
	The visible mean velocity is related to the mean velocity of the
	full system through the marginalization
	\begin{equation}
		\tilde{\nu}_{i} (\xt) \equiv \int \nu_{i} (\xh,\xt) p_{\mathrm{s}}
		(\xh|\xt) \, \mathrm{d}\xh \,,
	\end{equation}
	where the index $i$ is allowed to take values of the visible coordinates and
	$p_{\mathrm{s}}(\xh|\xt)$ is the stationary probability to be in the
	hidden state $\xh$ provided the state $\xt$ is observed.

	\subsection{Discrete Systems}
	In the same spirit it is possible to define the apparent entropy production
	if the system of interest is characterized by a discrete and finite set of
	states. As in the continuous case, we assume that each microstate is
	characterized by a visible label $\nt$ and a hidden internal state $\nb$.
	Microstates with the same visible label are grouped into mesostates.
	Transitions between the microstates take place with rates $w((\nt, \nb)
	\rightarrow (\nt',\nb'))$ and we denote the stationary distribution by
	$p_{\mathrm{s}}(\nt,\nb)$. The observable stationary distribution of the
	visible states is given by 
	\begin{equation}
		p_{\mathrm{s}}(\nt)	= \sum_{\nb} p_{\mathrm{s}}(\nt,\nb)\,.   
	\end{equation}

	Due to the fact that not every transition is observable and
	that there is an ambiguity which transition takes place even if a
	transition is observed, it is in general impossible  to determine the
	amount of produced entropy if one only observes a trajectory of the visible
	states. The only exception is a clear time-scale separation between the
	transitions within one mesostate and transitions between two different
	mesostates. In this case the dynamics can be described as a Markovian jump
	process on the level of the visible states and the total entropy production
	can be calculated as usual using the transition rates of this effective process.

	Our main focus lies on the case where such a time-scale separation is not
	present. Even though the visible dynamics is not Markovian one can still
	define effective transition rates $\wt(\nt \rightarrow \nt')$ that describe
	how often a certain transition will occur on average per time unit provided
	that the system is initially in state $\nt$. For Markovian dynamics
	this definition would lead to the actual transition rates.

	If the microscopic transition rates are known, the effective rates can be
	obtained by summation over all microscopic transitions that lead to a
	specific visible transition
	\begin{equation}
		\wt(\nt \rightarrow \nt') \equiv \frac{1}{\ps(\nt)}  \sum_{\nb,\nb'}
		 w((\nt,\nb) \rightarrow (\nt',\nb'))\ps(\nt,\nb)\,.
	\end{equation}
	These effective rates play the same role for the apparent entropy
	production in the discrete case as the mean velocity of the visible degree
	of freedom does in the continuous case. In the continuum limit, the mean
	velocity can indeed be obtained from the effective rates. In this limit, the
	effective rates are related to the local mean velocity conditioned on the
	visible state $\nut(\xt)$ in the same way as the microscopic rates are
	related to the unconditioned local mean velocity $\nuv(\vec{x})$. 

	For discrete systems, the true entropy production along a trajectory is
	defined as~\cite{seifert_stochastic_2012}
	\begin{equation}
		\stot = \sum_{n \rightarrow n'} \ln \frac{w(n \rightarrow n')}{w(n'
		\rightarrow n)} + \ln \frac{p_{\mathrm{s}}(n_{\mathrm{i}}) }
		{p_{\mathrm{s}}(n_{\mathrm{f}})}\,,
		\label{eq:stot_disc}
	\end{equation}
	where the sum runs over all jumps $n \rightarrow n'$ of the trajectory that
	begins in state $n_{\mathrm{i}}$ and ends in state $n_{\mathrm{f}}$. If
	there are hidden degrees of freedom, we define the apparent entropy
	production by replacing all quantities in~\eref{eq:stot_disc} with
	their observable counterparts, leading to
	\begin{equation}
		\tstot \equiv \sum_{\nt \rightarrow \nt'}
		\ln \frac{\wt(\nt \rightarrow \nt')}{\wt(\nt' \rightarrow \nt)} +
		\ln \frac{\ps(\nt_{\mathrm{i}} )}{\ps(\nt_{\mathrm{f}}) }\,.
		\label{eq:def_app_disc}
	\end{equation}

	The long time limit of the distribution of apparent entropy production is
	captured by the rate function
	\begin{equation}
		h(u) = 	\lim_{t \rightarrow \infty} \frac{1}{t} \ln p(\tstot=u t,t).  	
	\end{equation} 
	Since the last term in \eref{eq:def_app_disc} does not increase with time,
	it can be neglected for the calculation of the rate function.

	We also assign a true affinity $\Ac_{\Cc} $ and an apparent affinity
	$\At_{\Cc} $ to every
	closed cycle $\Cc$ in state space as the sum of all increments of true or
	apparent entropy production along the links of the cycle, i.e.
	\begin{equation}
		\Ac_{\Cc}  \equiv \sum_{n \rightarrow n' \in \Cc} \ln \frac{w(n \rightarrow
		n')}{w(n' \rightarrow n)} \quad \text{and} \quad \At_{\Cc}  \equiv  \sum_{ \nt
		\rightarrow \nt' \in \tilde{\Cc}} \ln \frac{\wt(\nt \rightarrow
		\nt')}{\wt(\nt' \rightarrow \nt) } \,,
		\label{eq:affinities}
	\end{equation}
	where $\tilde{\Cc}$ denotes the observable cycle corresponding to the cycle
	$\Cc$.

	Even though the apparent entropy production is in general different from the
	true entropy production, we will nevertheless show that it
	approximately obeys a modified FT in the long time limit
	if certain conditions are met.

	\section{Bipartite lattice as a model system} 
	\label{sec:bipartite_grid}

	\subsection{General setup}
\begin{figure}
		\centering
			\includegraphics{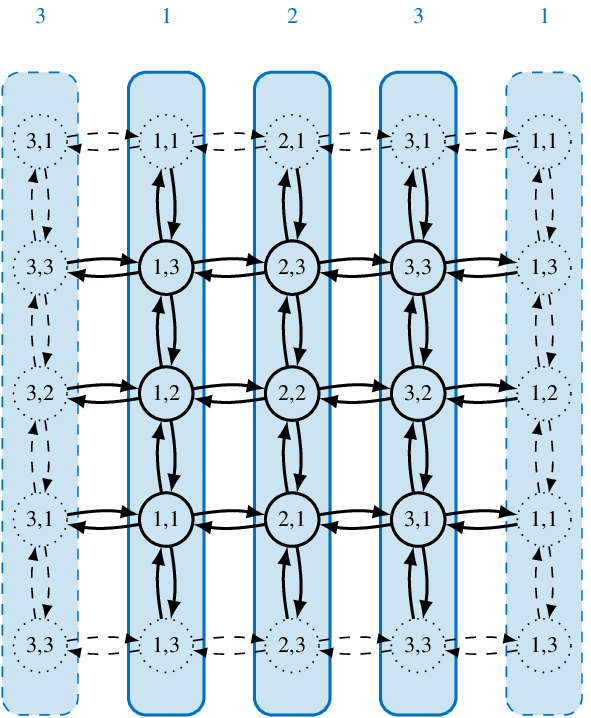}
		\hspace{0.1\textwidth}
			\includegraphics{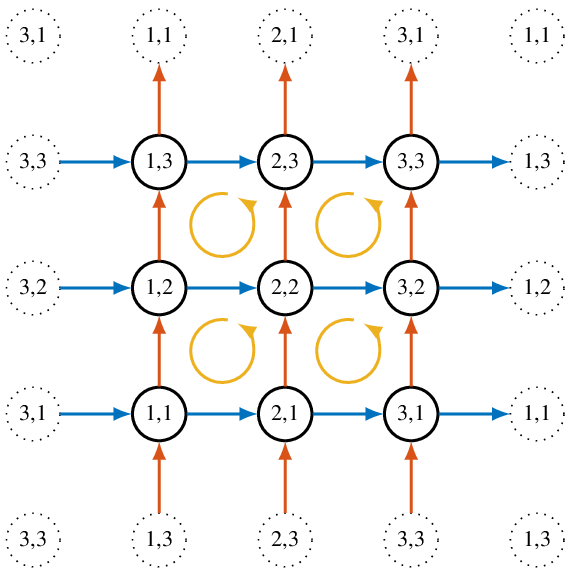}
			\caption{Visualization of the bipartite lattice for $N=3$. Each state
				is labeled by a visible and a hidden coordinate. States
				with the same visible coordinate are grouped together
				into visible states. The graph on the right hand side
				shows the fundamental cycles of the network. These can
				be divided into three groups depending on their
				affinities and effective affinities.}
		\label{fig:grid}
	\end{figure}

	We now present our system of interest. It consists of a two
	dimensional lattice of states $(n_{1},n_{2})$    with $N$ states in each
	dimension, as shown in \fref{fig:grid}.  Transitions are possible between
	neighboring states with periodic boundary conditions.  We assume that only
	the coordinate $n_{1}$ is visible and microstates within a column are
	indistinguishable for the observer. 

	Such a grid has $N^2 +1$ fundamental cycles
	\cite{schnakenberg_network_1976}, which can be chosen as depicted
	in \fref{fig:grid}. As the colors indicate they can be split into three
	different categories. $N$ of the cycles consist only of jumps that increase
	the visible coordinate. We therefore refer to them as visible cycles. In
	the same way there are also $N$ hidden cycles that only have steps that
	increase the hidden coordinate. The other $(N-1)^2$ are cycles that do
	not contain jumps across the periodic boundary conditions.

	This system is, \textit{inter alia}, inspired by the system of two coupled
	colloidal particles investigated in~\cite{mehl_role_2012}, where the system
	is driven out of equilibrium by two driving forces each acting on
	one colloid.  To emulate this kind of driving in our discrete setup we
	demand that only cycles that correspond to a full rotation of a degree of
	freedom have a non vanishing affinity. Furthermore all affinities of the
	visible cycles should be equal to an affinity $\Ac_{1}$ and all affinities of the
	hidden cycles to another affinity $\Ac_{2}$. 

	A set of rates meeting these conditions can be obtained by
	the ansatz of rates satisfying local detailed balance
	\begin{eqnarray}
	\fl	w((n_{1},n_{2}) \rightarrow (n_{1}+1,n_{2})) &= w_{n_{1} }^{1} \exp \left(
			\left(
		\Ac_{1}/N  + V_{n_{1},n_{2}  } - V_{n_{1}+1,n_{2}  } \right)/2   \right) \\
	\fl	w(( n_{1}+1,n_{2} ) \rightarrow (n_{1},n_{2})) &= w_{n_{1} }^{1} \exp
		\left(\left(
	-\Ac_{1}/N  + V_{n_{1}+1,n_{2}  } - V_{n_{1},n_{2}  } \right)/2  \right) \\
	\fl	w(( n_{1}, n_{2} ) \rightarrow ( n_{1},n_{2}+1 )) &= w_{n_{2} }^{2} \exp
		\left(\left(
		\Ac_{2}/N + V_{n_{1},n_{2}  } -V_{n_{1},n_{2}+1  }     \right)/2\right) \\
	\fl	w(( n_{1},n_{2}+1 ) \rightarrow ( n_{1},n_{2} )) &= w_{n_{2} }^{2} \exp \left(
			\left(
	-\Ac_{2}/N + V_{n_{1},n_{2}+1  }  -V_{n_{1},n_{2}  } \right)/2  \right) \,,
	\end{eqnarray}
	where we introduce a potential energy $V_{n_{1}, n_{2}}$ for each state and a
	characteristic timescale $w^{i}_{n}$ for each transition.
	This choice of rates ensures thermodynamic consistency and thus connects
	the physical properties of the system to its stochastic
	description~\cite{seifert_stochastic_2012}.
	If one chooses $w^{i} \propto N^{2}$ and the potential landscape as used in
	\cite{mehl_role_2012}  as $V$, the discrete system becomes equivalent to the
	continuous system in the limit $N \rightarrow \infty$.
	
	 \begin{figure}[htb]
		\centering
		\begin{minipage}{0.35\textwidth}
			\centering
			\includegraphics[width=\textwidth]{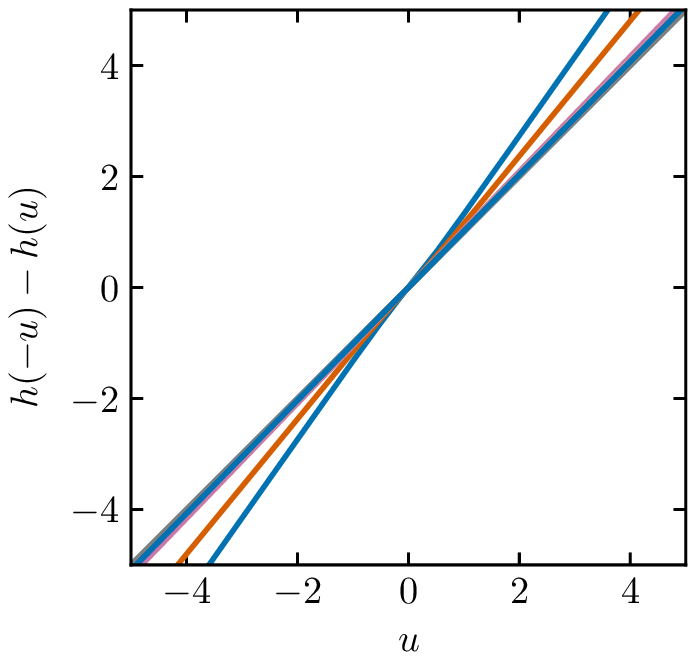}
		\end{minipage}
		\hspace{0.05\textwidth}
		\begin{minipage}{0.55\textwidth}
			\centering
			\includegraphics[width=\textwidth]{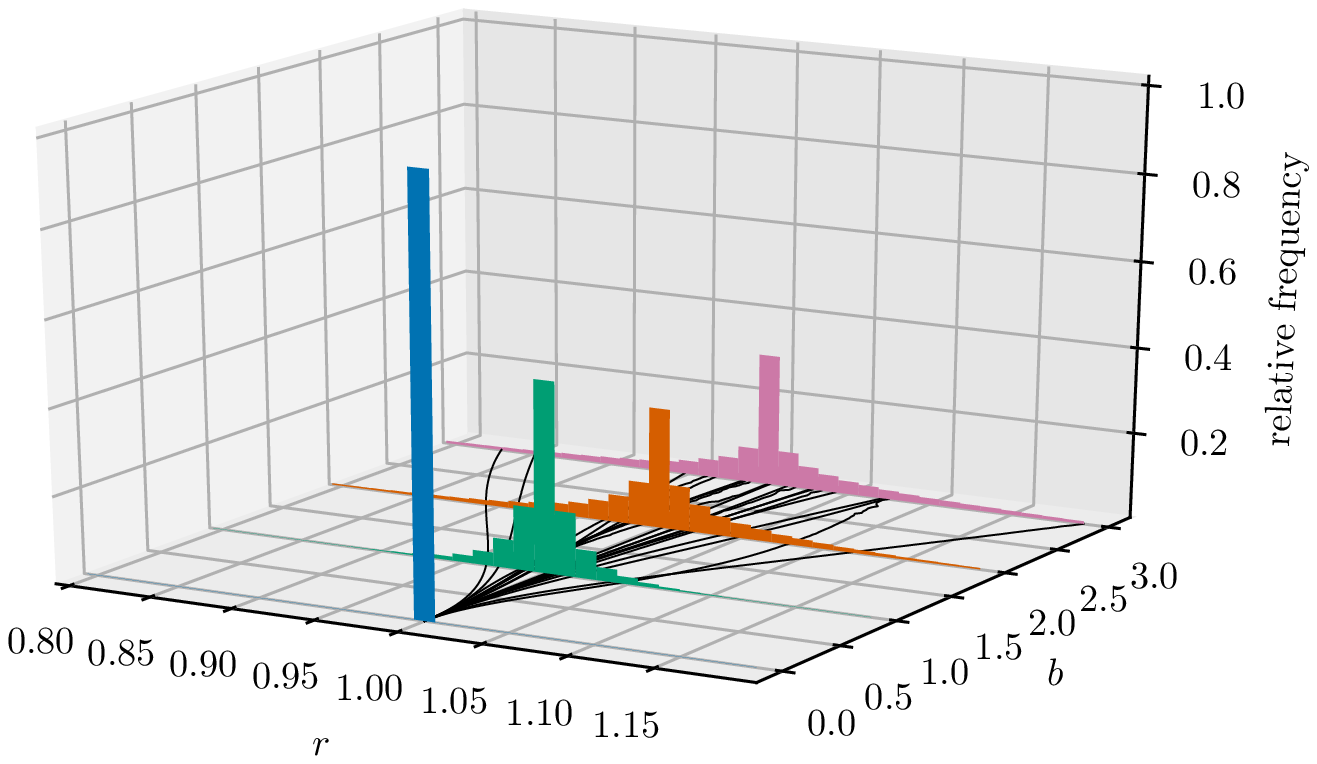}
		\end{minipage}
		\caption{Fluctuation relations for ten randomly generated sets of rates
			(left) for parameters $\Ac_{1}=\Ac_{2}=b=2$ and $N=3$. In most cases the
			antisymmetric part of the generating function is approximately
			linear. This is also evident from the histograms on the right. They
			show the distribution of the ratio $r$ \eref{eq:ratio} between
			slopes at zero and infinity of $10^5$ randomly drawn sets of rates at
			different amplitudes of the coupling $b$. The curves on the bottom
			show this ratio for 20 different potential landscapes drawn from a
			Gaussian distribution with variance 1 that are scaled by the factor
			$b$.}
		\label{fig:fluk_statistics}
	\end{figure}
	\subsection{Numerical results for random transition rates}
	We now present our first main result. For a small number of
	states the large deviation function for the apparent entropy production in
	a bipartite lattice obeys a modified FT for generic rates.
	Extensive numerical evidence shows that the antisymmetric part
	of the rate function typically does not deviate significantly from a linear
	behavior if the timescales $w_{n}^{i}  $ and the potential are drawn at random.
	The ensemble of random variations of the system is defined as follows: The potential energy
	$V_{n_{1},n_{2}}$ for each state is drawn from a Gaussian distribution with
	mean zero and standard deviation $b$. The timescales are chosen as
	\begin{equation}
		w_{n}^{i} = \exp(\psi^{i}_{n} - \bar{\psi^{i}} ),
	\end{equation}
	where the elements $\psi^{i}_{n} $ are drawn from a uniform
	distribution between 0 and 2 and $\bar{\psi^{i}} \equiv \sum_{n}
	\psi^{i}_{n}/N$ denotes the average value
	of the elements.
	The rates drawn from this ensemble are then used to calculate first the
	effective rates for this system and then the rate function $h(u)$ of the
	apparent entropy production (for details on the procedure see
	\sref{sec:large_deviations}). A few representative results of the
	antisymmetric part of the rate function $h(-u) -h(u)$ for $N=3$ are shown on the left
	panel of \fref{fig:fluk_statistics}. They are evidently linear, which means
	that a modified FT is valid for the distribution of
	apparent entropy production in these cases. To affirm this finding with
	better statistics we perform this calculation repeatedly for different
	values of $b$ and check the result for linearity. This is done by
	calculating the ratio 
	\begin{equation}
		r\equiv a_{0}/a_{\infty}
		\label{eq:ratio}
	\end{equation}	
	 of the slope at $u=0$, $a_{0} \equiv -2 h'(0)$, and the asymptotic
	 slope $a_{\infty} \equiv \lim_{u \rightarrow \infty} (h(-u) -h(u))/u $.
	 The latter is well defined and can be calculated directly from the
	 (effective) rates as will be explained in
	 \sref{sec:bounds_and_approximations}.

	The right panel of \fref{fig:fluk_statistics} shows histograms of the
	quantity $r$ demonstrating the approximate validity of an FT for different
	coupling strengths calculated  from $10^5$ numerical experiments for each
	different value of $b$. The affinities of the observable and hidden cycles
	where chosen as $\Ac_{1}=\Ac_{2}=1$. If there is no potential present
	($b=0$), the ratio is exactly 1 and the antisymmetric part of the rate
	function is linear in all cases  since the two degrees of freedom are
	independent and the effective rates are identical to the jump rates of the
	visible degree of freedom. Interestingly, this behavior changes only little
	even if the coupling becomes strong. For example, in the case $b=3$, the
	average potential difference of one jump alone exceeds the affinity of one
	full rotation of a degree of freedom but nevertheless a modified
	FT is still valid for this parameter combination in most
	cases.

	\section{Bounds and approximations}
	\label{sec:bounds_and_approximations}
	The numerical results for the rate function for apparent entropy production
	in the previous section have shown that a modified FT
	generically holds in very good approximation. In this section we will
	focus on the  underlying structure that leads to this behavior.

	\subsection{Large deviations of integrated currents}
	\label{sec:large_deviations}
	First we introduce the elements from large deviation theory needed for the
	scope of our calculations. For a more complete
	introduction, see, e.g.\,,~\cite{touchette_large_2009}.

	The quantities of interest in this paper are distributions of integrated
	currents. In a stochastical system with discrete states $\{n\}$ integrated
	currents are functionals of the trajectory $n(t')$ 
	where we assign an increment $d_{n,n'}=-d_{n',n}$ to each
	transition $n \rightarrow n'$ and sum over all transitions that have
	occurred
	\begin{equation}
		X[n(\cdot)] =  \sum_{n \rightarrow n'} d_{n,n'}\,.
		\label{eq:int_cur_disc}
	\end{equation}

	In the continuous limit this sum becomes an integral along the trajectory
	\begin{equation}
		X[\xv(\cdot)] = \int_{0}^{t} \vec{d}(\xv(t')) \cdot \dot{\xv}(t') \, \mathrm{d}t' \,.
	\end{equation}
	The apparent entropy production is an integrated current with the choice
	$d_{\nt, \nt'} = \ln \wt(\nt \rightarrow \nt')/\wt(\nt' \rightarrow \nt)$ in the
	discrete and $\vec{d}(\xv) = \nut(\xv)$ in the continuous case.

	The results derived in this paper are essentially relations and statements
	about the long time behavior of the probability distributions  $p(X,t)$ of
	integrated currents. It can be shown that for long enough times these
	probabilities will decay exponentially, thus motivating the definition of a
	rate function~\cite{touchette_large_2009}
	\begin{equation}
		h(u)  \equiv  - \lim_{t \rightarrow \infty} \frac{1}{t} p(X =
		ut, t)  
		\label{eq:def_rate_function}
	\end{equation}
	that characterizes the speed of this decay.

	The rate function can be obtained
	without the explicit knowledge of the distribution through
	Legendre-Fenchel-transformation of the
	rescaled cumulant generating function
	\begin{equation}
		\alpha(\lambda) \equiv \lim_{t \rightarrow \infty} \frac{1}{t} \ln
		\left\langle  e^{\lambda X} \right\rangle
		\label{eq:def_gen_funk}
	\end{equation}
	as
	\begin{equation}
		h(u) = \sup_{\lambda} \left[ u\lambda - \alpha(\lambda) \right]\,.
	\end{equation}

	In general, the generating function can be obtained by determining the time
	evolution of the moment generating function conditioned on the final state
	of the trajectory
	\begin{equation}
		g(\lambda, \xv, t) \equiv 	\left\langle e^{\lambda X}
		\right\rangle_{\xv}\,, 
	\end{equation}
	where the possibly discrete index $\xv$ indicates that for the expectation
	value only those trajectories are taken into account that end in state
	$\xv$.  This time evolution is generated by the so called tilted master
	operator $\Lc(\lambda)$ leading to
	\begin{equation}
		\partial_{t} g(\lambda, \xv, t) = \Lc(\lambda)g(\lambda, \xv, t) \,.
	\end{equation}
	In the case of an integrated current defined on a discrete set of states as
	in \eref{eq:int_cur_disc} the explicit form of $\Lc(\lambda)$ reads
	\begin{equation}
		\Lc(\lambda) = w(n \rightarrow n') \exp(d_{n',n} \lambda) - \delta_{n,n'}
		\sum_{n'} w(n \rightarrow n')\,.
	\end{equation}
	It can be shown that the generating function $\alpha(\lambda)$ is the Perron-Frobenius
	eigenvalue of the tilted operator $\Lc(\lambda)$. We can therefore
	calculate the rate function by diagonalizing this operator and performing a
	Legendre-Fenchel-transformation of the eigenvalue with the largest real
	part.

	This paper is concerned with the question whether the distribution of
	certain integrated currents obey a FT of the form $\ln
	(p(X,t)/p(-X,t)) = \lt X$ in the long time limit, where $\lt$ corresponds
	to the slope $a$ of the modified FT~\eref{eq:fluc_theorem}.  In terms of the rate
	function and the generating function, such a relation is expressed by the two
	equivalent relations 
	\begin{equation}
		\alpha(\lambda) = \alpha(-\lt -\lambda) \quad
		\Leftrightarrow \quad h(-u) - h(u) = \lt u \,.
		\label{eq:symmetry}
	\end{equation}
	In words, the rate function obeys a modified FT if and only
	if the generating function is symmetric around a value corresponding to the
	negative slope appearing in the FT.

	One important aspect to note at this point is that since we know that
	$\alpha(\lambda)$ is convex, the Legendre-Fenchel-Transformation is a one
	to one mapping between individual points of the two functions.
	Therefore~\eref{eq:symmetry} is also valid locally in the sense that, if
	there exists some, not necessarily connected, region in the $\lambda$
	domain in which the left hand side of~\eref{eq:symmetry} holds, there must
	also exist a region in $u$ where the right hand side holds.
	
	\subsection{Asymptotic behavior of the rate function}
	\label{ssec:asymptotics}
	In previous work~\cite{pietzonka_universal_2016}, it was
	shown that the generating function of any integrated current in a discrete
	network will approach an exponential function from above as its argument goes to
	infinity. 
	For the present problem this asymptotic bound can be constructed as
	follows.
	If we choose an arbitrary closed path $\Cc$ on the
	network with length $N_{\Cc}$, which we denote as 
	\begin{equation}
		\Cc  \equiv \left[ n(1) \rightarrow n(2) \rightarrow \cdots \rightarrow
		n(N_{\Cc}) \rightarrow n(1)   \right] \,,
	\end{equation}
	we can assign to it the geometric mean of transition rates along the path
	\begin{equation}
		\gamma_{\Cc} \equiv \left( \prod_{i=1}^{N_{\Cc}-1} w(n(i) \rightarrow
		n(i+1))   \right)^{1/N_{\Cc} } 
	\end{equation}
	as well as the number $\mathfrak{n}_{\Cc}$  of visible cycles completed by
	the path $\Cc$.  It was shown in~\cite{pietzonka_universal_2016} that each path implies a lower bound on
	the generating function for the apparent entropy production of the form
	\begin{equation}
		\alpha(\lambda) \geq f(\lambda,\Cc) \equiv  \gamma_{\Cc} \exp
		\left(\frac{\mathfrak{n}_{\Cc} }{N_{\Cc} } \At \lambda \right) -
		\max_{n} r_{n} \,,
		\label{eq:asympt_bound}
	\end{equation}
	where $\At$ is the apparent affinity from \eref{eq:affinities} of the
	unique cycle in the visible state space, and where $r_{n} $ denotes the
	exit rate of the state $n$.
	
	For large $\lambda$, the strongest bound is achieved by the path $\Cc_{+}$ that
	has the largest ratio of completed visible cycles $\mathfrak{n}_{\Cc}$ to
	number of jumps $N_{\Cc}$. Obviously this is a cycle that only jumps in
	positive $n_{1}$ direction while maintaining the state of  the
	hidden degree of freedom (shown in blue in \fref{fig:grid}), since any
	additional step in the hidden direction would increase the length of the
	cycle while maintaining the same winding number and would therefore lead to
	a lower ratio.
	Among those cycles the one with the highest geometric mean of the rates
	leads to the strongest bound.

	For $\lambda \rightarrow - \infty$ the optimal cycle is the same one but
	traversed in opposite direction. We denote this path by $\Cc_{-}$.
	The bound induced by $\Cc_{-}$ is related to the one induced by
	$\Cc_{+}$ through the relation
	\begin{equation}
	f(\lambda, \Cc_{-}) = f \left(-\frac{\Ac_{1} }{\At} -\lambda,\Cc_{+}
	\right)\,,
	\label{eq:bound_symmetry}
	\end{equation}
	as can be shown using the relation of the geometric mean in forward and
	backward direction with the affinity of a visible cycle
	\begin{equation}
		\frac{\gamma_{\Cc_{+} }}{\gamma_{\Cc_{-}  }  } = \exp \left(
		\frac{\Ac_{1} }{N_{\Cc} }  \right)\,.
	\end{equation}

	The numerical evidence presented
	in~\cite{pietzonka_universal_2016} indicates that this bound becomes
	tight in the limit $\lambda \rightarrow \pm \infty$ in the sense that the
	ratio of the generating function and the bound converges to 1. As a
	consequence,~\eref{eq:bound_symmetry} implies that the tails of the
	generating function are symmetric around $\lambda = -\Ac_{1}/(2\At)$.
	Following \eref{eq:symmetry} the rate function hence
	obeys a modified FT of the form
	\begin{equation}
		h(-u) -h(u) = \frac{\Ac_{1} }{\At} u
		\label{eq:asym_slope}
	\end{equation}
	asymptotically for large $|u|$. 
	
	This relation also shows that, in principle, it is possible to infer the
	amount of true entropy produced in one rotation of the visible
	degree of freedom from the statistics of the extremes of the apparent entropy
	production, since these are generated by trajectories for which the
	hidden degree of freedom is not changing.
	The only knowledge about the system that is a priori necessary for such an
	inference is that the visible degree of freedom has indeed a cyclic
	structure, with which the affinity $\Ac_{1}$ can be associated, and that
	the coupling strength to the hidden degree of freedom is finite. Further
	knowledge, e.g. about the transition rates or the number of visible and
	hidden states, is not required.

	\subsection{Symmetric approximation of the generating function}
	\label{ssec:sym_approx}
	The numerical evidence presented in \sref{sec:bipartite_grid} indicates
	that the linear behavior of the antisymmetric part of the rate function is
	not limited to the extremes of the distribution but is also present for
	smaller arguments of the rate function. The goal of this section is to
	derive a symmetric approximation on the characteristic polynomial of the
	tilted master operator defined as
	\begin{equation}
		\chi(\lambda,y) \equiv \det(\Lc(\lambda)-y) = \sum_{\pi \in S_{N^{2}}}
		\sgn(\pi) \prod_{n}\left[ \Lc_{n,\pi(n)}(\lambda) - y
		\delta_{n,\pi(n)} \right]\,,
		\label{eq:charpol}
	\end{equation}
	where the sum runs over the set $S_{N^{2}}$ of all possible permutations $\pi$ of state space.
	Since the generating function is the largest eigenvalue of $\Lc(\lambda)$
	and hence the largest root of $\chi(\lambda, y)$ in $y$ for any given
	value of $\lambda$, the validity of such an approximation would imply a
	modified FT.

	Because the fundamental cycles of the bipartite grid split into
	three categories  depending on their combination of effective and actual
	affinity as described in \sref{sec:bipartite_grid}, the sum over
	permutations can be expressed in the form (see appendix)
	\begin{equation}
		\fl \chi(\lambda, y) = \sum_{\pi \in S_{N^{2}}} \sgn(\pi)
		f_{\pi}(y)  \cosh
		\left[\mathfrak{n}_{\pi} \left(\tilde{\mathcal{A}}\lambda +
		\mathcal{A}_{1}/2 \right) + \mathfrak{m}_{\pi} \mathcal{A}_{2}/2
		\right] \,. 
		\label{eq:charPol_final}
	\end{equation}
	The two integers $\nf$ and $\mf$ are winding numbers that count
	how often the permutation links states across the periodic boundary
	condition in the positive observable and hidden direction respectively.
	The polynomial $f_{\pi}(y)$ is the only part of each term that depends on
	$y$ and its degree is bounded from above by $N^2 - N(|\nf| + |\mf|)$.
	
\begin{figure}
		\centering
		\begin{minipage}{0.45\textwidth}
			\centering
			\includegraphics[scale=1]{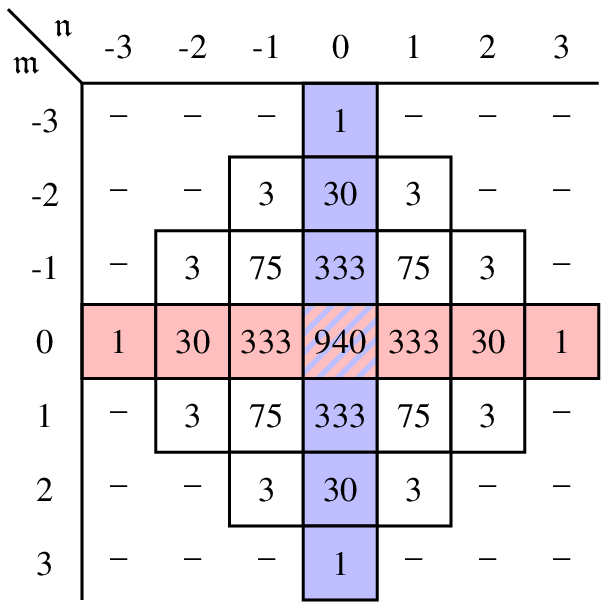}
		\end{minipage}
		\hspace{0.05\textwidth}
		\begin{minipage}{0.45\textwidth}
			\centering
			\includegraphics[scale=1]{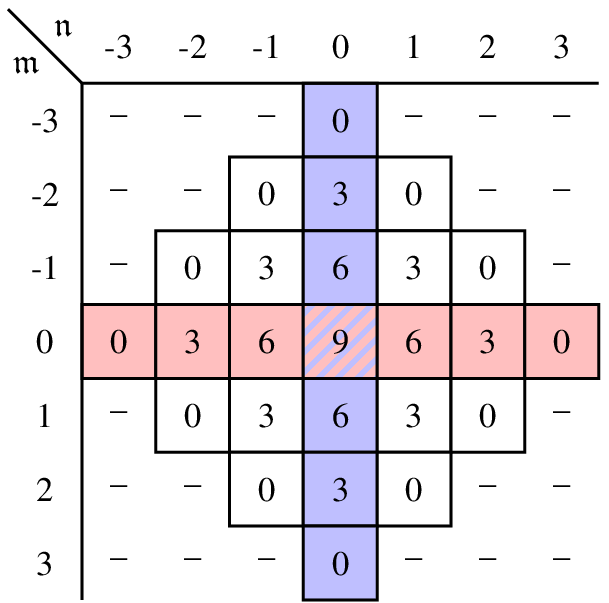}
		\end{minipage}
		\caption{Left: number of permutations with a specific
			combination of winding numbers for $N=3$. For most
			permutations either the visible winding number (blue) or
			the hidden winding number (red) vanishes.
			Right: The table shows the highest possible degree of the
			polynomial $f_{\pi}(y)$ corresponding to a permutation with given
		winding numbers $\nf$ and $\mf$. It is evident permutations with $\nf
	\cdot \mf \neq 0$ do not depend as strongly on $y$ as most other
permutations.}
		\label{fig:permutation_number}
	\end{figure}

	A symmetric approximation to the characteristic polynomial
	\begin{equation}
		\fl \chi^{\text{sym}}(\lambda, y) \equiv \sum_{\pi \in S_{N^{2}}| \nf
			\cdot \mf = 0} \!\!\!\!\!\!\!\! \sgn(\pi) f_{\pi}(y)  \cosh
		\left[\mathfrak{n}_{\pi} \left(\tilde{\mathcal{A}}\lambda +
		\mathcal{A}_{1}/2 \right) + \mathfrak{m}_{\pi} \mathcal{A}_{2}/2
		\right] 
		\label{eq:charPol_approx}
	\end{equation}	
	with the same
	center of symmetry as the asymptotes of the generating function can now be
	constructed by summing only over those terms where either the visible, $\nf$ or the
	hidden winding number $\mf$ vanishes. 
	That this procedure tends to be a good approximation at least for a low number of
	states $N$ can be understood simply by counting the remaining non-vanishing
	terms by their combination of winding numbers. The result is shown for
	$N=3$ in \fref{fig:permutation_number}. It turns out that skipping terms
	with $\nf \cdot \mf \neq 0$ omits only 324 of a total of 3720 
	terms.

	Furthermore it can be argued that terms with $\nf \cdot \mf
	\neq 0$ will not depend as strongly on $y$ and are therefore less
	significant for the position of the roots. The only part of the term
	depending on $y$ is the polynomial $f_\pi(y)$, whose degree is coupled
	to the winding numbers. The highest possible degree of $f_\pi (y)$ is listed
	in the right panel of \fref{fig:permutation_number} depending on the values of the two winding numbers.

\begin{figure}
		\centering
		
		\begin{minipage}{0.30\textwidth}
			\centering
			\includegraphics[width=\textwidth]{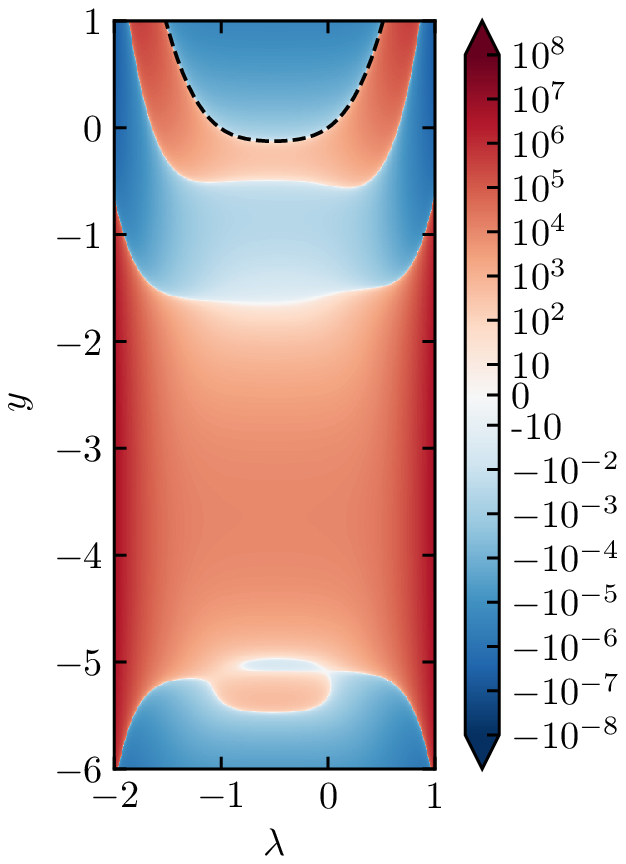}
		\end{minipage}
		\hspace{0.2\textwidth}
		\begin{minipage}{0.30\textwidth}
			\centering
			\includegraphics[width=\textwidth]{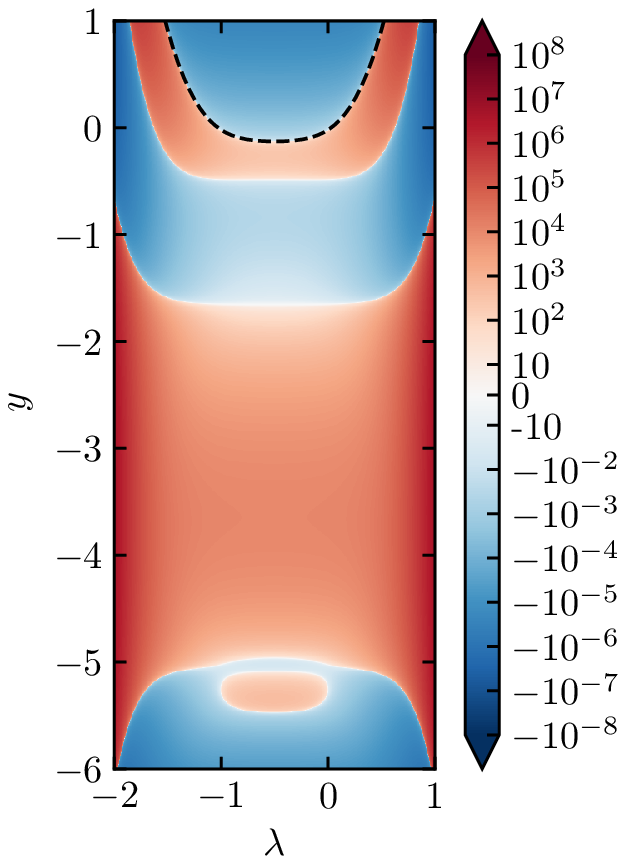}
		\end{minipage}
		\caption{Logarithmic plot of the characteristic polynomial
			$\chi(\lambda, y)$ (left) and
			the symmetric approximation $\chi^{\text{sym}}(\lambda, y)$ (right) obtained by only summing over terms
			with at least one vanishing winding number. This is a good
			approximation in the region where the largest eigenvalue (indicated as dashed line) lies, which explains the
			approximate symmetry of the generating function.}
		\label{fig:approximation}
	\end{figure}

	An example for a typical characteristic polynomial and the corresponding
	symmetric approximation is shown in \fref{fig:approximation}. It is
	evident that this approximation not only captures the behavior of the
	largest eigenvalue but also describes the behavior of the subdominant eigenvalues
	quite well.

	\subsection{Deviations from the fluctuation theorem}
	\label{sub:deviations}

	While the omission of terms in the characteristic polynomial that are not
	symmetric under the replacement $\lambda \rightarrow -\Ac_{1}/\At
	-\lambda$ leads to an approximation that works rather well in most cases, we
	can also identify those rare cases where the generating function is
	pronouncedly asymmetric. In these instances terms with $\nf \cdot \mf
	\neq 0$ must make a significant contribution to the characteristic polynomial. 
	This happens if the rates are large along a cyclic path for which both winding
	numbers do not vanish. Moreover, rates that lead away from this
	path have to be small while rates pointing towards the path are large.
	These conditions are met for example if we choose the potential as 
	\begin{equation}
		V_{n_{1},n_{2}  } = V_{0}  \left( \begin{array}{ccc}
				0 & 0 & 1 \\
				1 & 0 & 0 \\
				0 & 1 & 0 
		\end{array} \right). 
	\end{equation}
	If $V_{0}$ is sufficiently large, this potential will force the system to
	jump only between states with vanishing potential while jumps onto the
	potential barriers are suppressed. 

\begin{figure}
		\centering
		\begin{minipage}{0.35\textwidth}
			\includegraphics[scale=0.5]{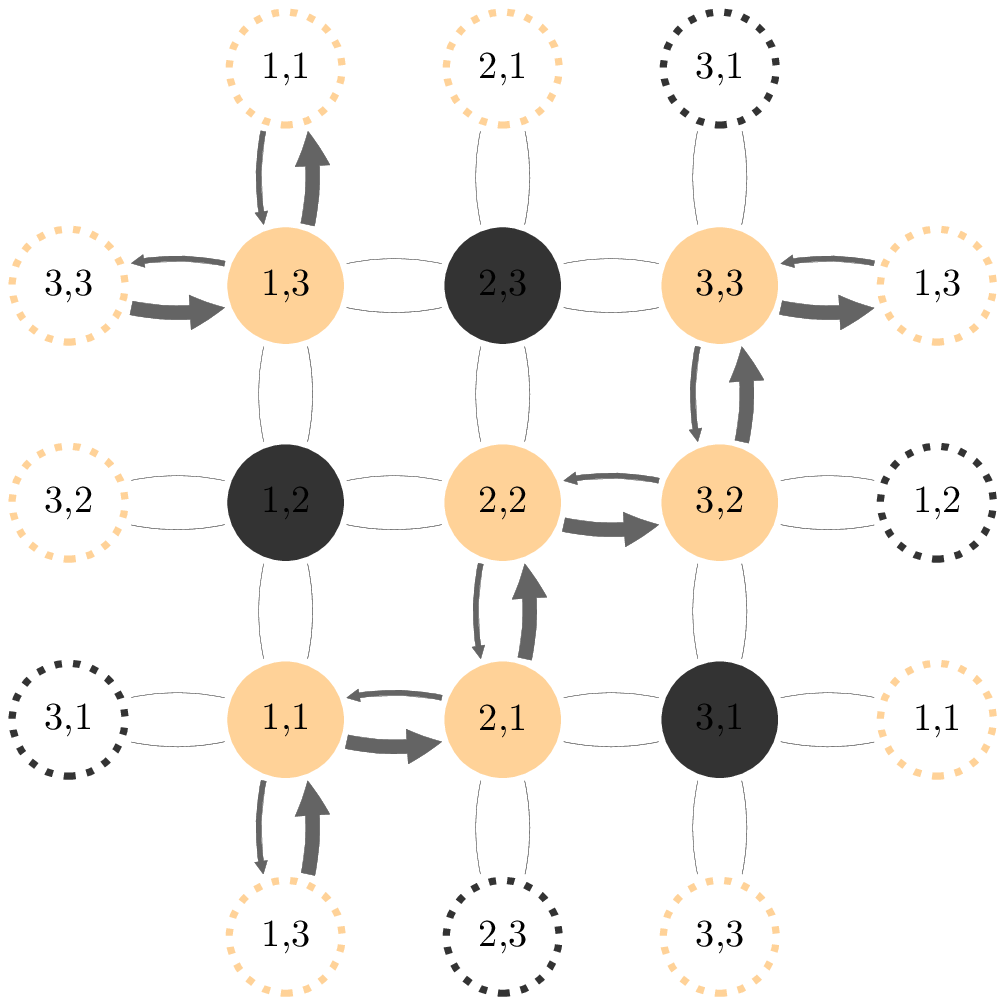}
		\end{minipage}
		\hspace{0.05\textwidth}
		\begin{minipage}{0.55\textwidth}
			\includegraphics[width=\textwidth]{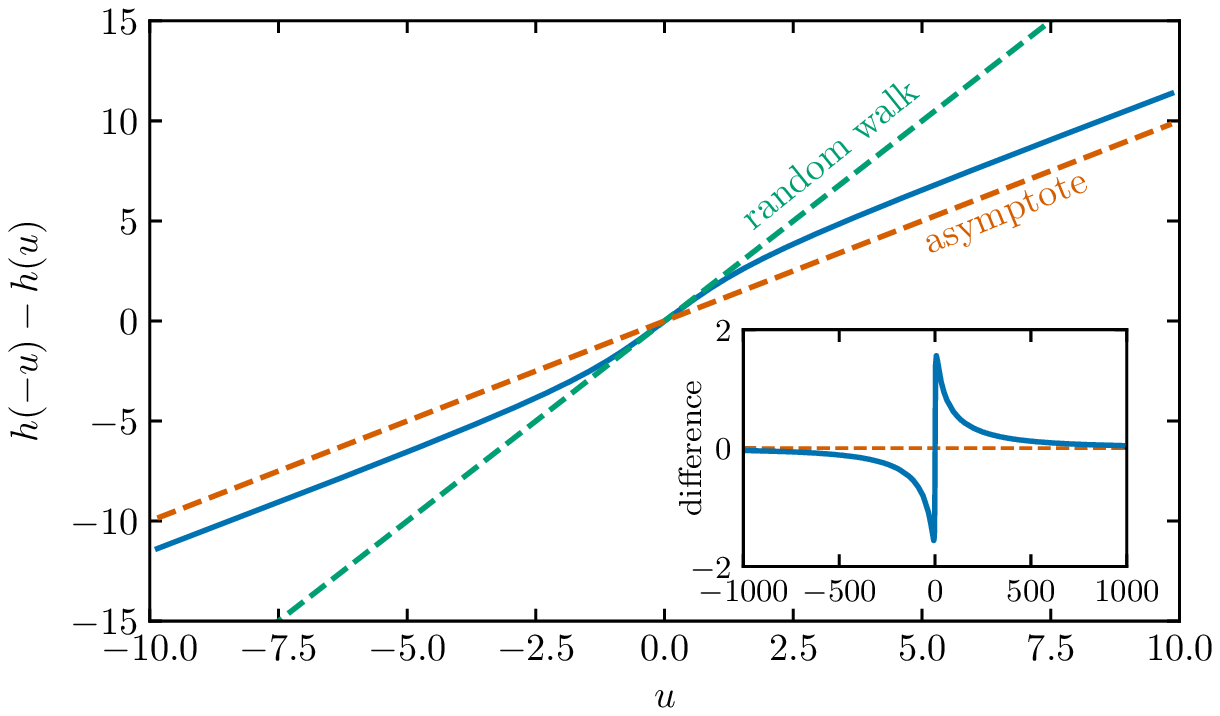}
		\end{minipage}
		\caption{Stationary distribution and antisymmetric part of the
		rate function for a network with rates designed to break the modified
		FT. The states are colored according to their
	respective stationary distribution (lighter means higher $\ps$). The width
	of the arrows are scaled by the likelihood of observing a specific
	transition $w(\nv \rightarrow \nv')\ps(\nv)$. The affinities where set to
	$\Ac_{1}=\Ac_{2}=3$ and the height of the potential barrier to $V_{0}=10$.
	The antisymmetric part of the rate function shows a pronounced non-linear
	behavior close to the origin and reaches a linear asymptote. The inset
	shows the difference to the asymptote over a larger interval of $u$. The
slope at the origin is described by the appropriately rescaled FT for the
asymmetric random walk.}
		\label{fig:zigzag}
	\end{figure}

	As a result of the lack of symmetry in the generating function, the
	antisymmetric part of the rate function shows a pronounced non-linear
	behavior for small values of $|u|$ while it still approaches an asymptotic
	slope for large $|u|$ as a consequence of the asymptotic
	relation~\eref{eq:asym_slope}. This is shown in \fref{fig:zigzag} where the
	affinities of the two degrees of freedom were chosen to be equal.

	In this specific case the system will most likely perform an asymmetric
	random walk along the potential ``valley". For this kind of dynamics, the
	generating function $\alpha_{\text{RW}}(\lambda)$  and the rate function
	$h_{\text{RW}}(u)$ for the total entropy production are known and obey the
	usual Gallavotti-Cohen-symmetry~\cite{lebowitz_gallavotti-cohen-type_1999} and the
	FT. Since we are considering not the total but the apparent entropy
	production these results have to be rescaled by the ratio of the total
	entropy production to the apparent entropy production per rotation, i.e.
	\begin{eqnarray}
		\alpha(\lambda) = \alpha_{\text{RW}} \left( \frac{\At}{\Ac_{1} +
		\Ac_{2} } \lambda \right)\\
		h(u) = h_{\text{RW}}  \left( \frac{\Ac_{1} +\Ac_{2}  }{\At} u
		\right)\,.
	\end{eqnarray}
	We therefore expect the rate function to obey a modified FT with
	slope $(\Ac_{1} +\Ac_{2})/\At$ for small $|u|$ if the potential barriers
	are large as shown in~\fref{fig:zigzag}.

	On the other hand the asymptotic bound is determined by those
	periodic trajectories that minimize the apparent entropy production
	per jump. Clearly those are the trajectories that always jump in
	one direction along the visible axis. They produce $\Ac_{1}$ of total
	entropy per turn, which leads with the same arguments as above to
	the slope $\Ac_{1}/\At$.

	\section{Continuous case}
	\label{sec:continuous_case}

	So far, we have explained the generic emergence of a modified FT for the
	bipartite lattice. Transferring this result to a system with continuous
	degrees of freedom governed by a Langevin equation like the one used
	in~\cite{mehl_role_2012} is not trivial. First, the asymptotic bound used
	in the previous section to derive the asymptotic slope of the asymmetric
	part of the rate function is not valid for continuous systems since the
	limit $\lambda \rightarrow \infty$ does not commute with the limit $N
	\rightarrow \infty$. When the number of states increases, the region in
	which the asymptotic bound is a good approximation moves outwards leading
	to quadratic (rather than exponential) tails of the generating function in
	the case of a continuous state space. 

	\subsection{Asymptotic bound}
	\label{sub:Asymptotic bound}
	
	In this section, we will derive a bound on the generating function similar
	to the auxiliary bound introduced in~\cite{tsobgninyawo_large_2016} for a one
	dimensional system
	that will become tight up to a constant for large $|\lambda|$ and is also
	symmetric. It can therefore replace the asymptotic bound in the derivation
	of the asymptotic slope of the antisymmetric part of the rate function.
	Interestingly, it will turn out that the center of symmetry of this new
	bound is the same as for the one used in the discrete case.  This means
	that the asymptotic slope survives the continuum limit despite the fact
	that the bound which it was originally based on does not.

	We consider  two dimensional dynamics governed by the coupled Langevin
	equations
	\begin{eqnarray}
		\dot{x}_{1}(t) = f_{1} - \partial_{x_1 }V(x_{1},x_{2}) +
		\xi_{1} (t) \\
		\dot{x}_{2}(t)= f_{2} - \partial_{x_2 }V(x_{1},x_{2}) +
		\xi_{2}(t)      
	\end{eqnarray}
	with periodic boundary conditions identifying $x_{i}$ and $x_{i} + 2 \pi$ and the
	random forces $\xi_{i}(t)$ with 
	\begin{equation}
		\left\langle \xi_{i}(t)\xi_{j}(t')   \right\rangle =
		2\delta_{i, j} \,\delta(t-t') \,.
	\end{equation}

	In the long time limit, the apparent entropy production depends only on the
	distance $\Delta x_{1}$  traveled by the first degree of freedom, since
	\begin{equation}
		\tstot = \int_{0}^{t}   \tilde{\nu}_{1} (x_{1}) \dot{x}_{1}(t') \, \dtp \approx \frac{\At}{2
		\pi} \Delta x_{1} \quad \text{with} \quad \At \equiv \int_{0}^{2 \pi}
		\tilde{\nu}_{1}(x_{1}) \, \mathrm{d}x_{1}\,.
		\label{eq:distance_entropy}
	\end{equation}
	up to finite boundary terms.
	Therefore, we calculate first the generating function for the distance
	$\Delta x$, which we denote by $\alpha_{x}(\lambda_{x})$, and obtain the
	generating function for $\tstot$ through rescaling of this function.  
	We use the index $x$ to distinguish the respective functions and
	arguments for the
	traveled distance from the ones for apparent entropy production.
	The tilted operator for the distance was derived in~\cite{mehl_large_2008}
	and for a more general case in~\cite{chetrite_nonequilibrium_2013} and is given
	by
	\begin{equation}
		\mathcal{L}(\lax) = \Lfp + \Lone \lax +
		\lax^{2}   
	\end{equation}
	where $\Lfp$ denotes the Fokker-Planck operator
	\begin{equation}
		\Lfp = \sum_{i \in \{1,2\}} - \partial_{x_{i} }
		\left[F_{i}(\vec{x}) - \partial_{x_{i} } \right]   
	\end{equation}
	and 
	\begin{equation}
		\Lone = F_{1} (\vec{x}) - 2 \partial_{x_{1} } \,.
	\end{equation}
	From this form of the tilted operator it is obvious that its largest
	eigenvalue, the generating function, does not increase exponentially but
	quadratically $\alpha_{x} (\lax) = \lax^{2} + \mathcal{O}(\lax)$.

	In order to show that the generating function becomes symmetric, we use the
	general result from~\cite{maes_canonical_2008} (see
	also~\cite{barato_formal_2015, hoppenau_level_2016})
	for the rate function for the probability
	to observe a given empirical distribution
	\begin{equation}
		\varrho(\xv) \equiv \frac{1}{t} \int_{0}^{t} \delta(\xv(t')
		-\xv)\dtp
	\end{equation}
	and empirical currents
	\begin{equation}
		\muv(\xv) \equiv \frac{1}{t}  \int_{0}^{t} \dot{\xv}(t')
		\delta(\xv(t') -
		\xv) \dtp
	\end{equation}
	that is given by
	\begin{equation}
			I[\varrho(\cdot),\vec{\mu}(\cdot)] = \frac{1}{4} \int
			\mathrm{d} \vec{x} \; \varrho(\vec{x}) \left[
			\frac{\vec{\mu}(\vec{x}) - \vec{j}_{\varrho}(\vec{x})
			}{\varrho(\vec{x})}  \right]^{2} \,.
	\end{equation}
	Here $\vec{j}_{\varrho} (\xv)$ denotes the probability current according to
	\eref{eq:prob_current} if the distribution $p(\xv, t)$ is replaced by the
	empirical distribution $\varrho(\xv)$.
	Since the empirical current uniquely determines $\Delta x_{1}$, it is in
	principle possible to calculate the rate function $h_{x}(u_{x})$ by the
	contraction principle, as
	\begin{equation}
		h_{x}(u_{x}) = \min_{\varrho(\xv)} \min_{\muv(\xv) | u_{x}[\muv] =
		u_{x}  } I[\varrho(\xv),\muv(\xv)]\,,    
	\end{equation}
	where the minimization runs over all functions $\rho(\xv)$ and $\muv(\xv)$
	fulfilling the additional constraint $\nabla \muv(\xv)=0$
	that lead to the desired value of the integrated current.
	Although this minimization is not feasible for most systems, we can obtain
	an upper bound on the rate function if we insert a suitable ansatz and
	minimize with respect to the parameters therein.

	For our purpose it is sufficient to insert the arguably simplest form of
	empirical distribution and current, the constant functions
	\begin{equation}
		\varrho(\vec{x}) = \frac{1}{4 \pi^{2} } \quad \mathrm{and}
		\quad \vec{\mu}(\vec{x}) = \frac{1}{4 \pi^{2} } \left( \mu_{1}  \vec{e}_{1}+ \mu_{2}
		\vec{e}_{2}  \right)\,.
	\end{equation}
	To get a bound on the rate function for the apparent entropy
	production we have to minimize $I$ with respect to the free
	parameters $\mu_{1}$ and $\mu_{2}$ subject to the condition that
	the empirical current reproduces the desired value of $u_{x} $. This
	procedure fixes $\mu_{1}$, since    
	\begin{equation}
		u_{x}  = \frac{\Delta x_{1} }{t} = \frac{1}{4 \pi^{2} }  \int \mu_{1} \mathrm{d}\vec{x}=
		\mu_{1}  \,,
	\end{equation}
	leaving us with	
	\begin{eqnarray}
		\fl I[\varrho(\vec{x}),\vec{\mu}(\vec{x})] &= \frac{1}{4} \int
		\mathrm{d}\vec{x} \frac{1}{4 \pi^2} \left[ \left( u_x - f_{1}
			\right)^2 + 2 \left( u_{x} -f_{1}  \right)(\partial_{x_{1}} V(\vec{x}))
		+ \left( \partial_{x_{1}} V(\vec{x})  \right)^2 \right. \nonumber \\ &
		 \qquad \left. +( \mu_{2} -f_{2})^2 + 2 \left(\mu_{2} -f_{2}
				 \right) \left( \partial_{x_{2}}V(\vec{x})\right)  + \left(
		 \partial_{x_{2}} V(\vec{x}) \right)^2  \right] \nonumber\\
		\fl &=\frac{1}{4} \left[ (u_{x} -f_{1})^2 +
			\overline{(\partial_{x_{1}}V)^2}+ \overline{(\partial_{x_{2}}V)^2}
		\right] + \left( \mu_{2} -f_{2} \right)^2 /4\,,
	\end{eqnarray}	
	where we use the notation $\overline{A(\vec{x})} \equiv \int A(\vec{x})\,
	\mathrm{d}\vec{x}/(4\pi^2)$ to denote the average value of a function.
	This relation reaches a minimum if the last term vanishes. Minimization
	thus yields the upper bound on the rate function
	\begin{equation}
		h_{x} (u_{x} ) \leq \min_{\mu_{2} }
		I[\varrho(\vec{x}),\vec{\mu}(\vec{x})] = \frac{1}{4} \left[
			(u_{x} -f_{1})^2 + \overline{(\partial_{x_{1}}V)^2}+
		\overline{(\partial_{x_{2}}V)^2}  \right] 
	\end{equation}
	corresponding to a lower bound on the generating function
	\begin{equation}
		\alpha_{x} (\lambda_{x} ) = \max_{u_{x} } \left[ u_{x} \lambda_{x} \! - h_{x}
		(u_{x} ) \right] \geq
		 (\lambda_{x}  +f_{1})\lambda_{x} -\frac{1}{4} \left[
			\overline{(\partial_{x_{1}}V)^2}+ \overline{(\partial_{x_{2}}V)^2}
		\right].
	\end{equation}
	Since the bound and the generating function behave for large arguments like
	a quadratic function with the same curvature   and since they must not intersect,
	the bound can asymptotically differ only by a constant from the generating
	function
	\begin{equation}
		\alpha_{x} (\lambda_{x} ) =  (\lambda_{x} +f_{1})\lambda_{x} + \mathcal{O}(1) \,.
	\end{equation}
	Following the same arguments as in the discrete case, this scaling corresponds to a
	asymptotic linear behavior of the antisymmetric part of the rate function.
	\begin{equation}
		h_{x} (-u_{x} ) -h_{x} (u_{x} ) \sim f_{1} u_{x} \,.
	\end{equation}
	If we use the rescaling between the traveled distance and the apparent
	entropy production $u_{x} = 2 \pi u /\At$  , this relation is equivalent
	to the one found for discrete systems in~\eref{eq:asym_slope}.

	\subsection{Numerical case study}
	\label{sub:Numerical case study}
	While this expression shows that the
	asymptotics of the rate function obeys a modified FT also in the continuous case, this
	symmetry does not hold for smaller values of $|u|$ even approximately. The
	reason why the modified FT holds approximately for the
	discrete systems considered in~\sref{sec:bipartite_grid} lies in the small
	number of states and therefore small number of paths with non vanishing
	winding number in both directions. As the number of states increases in the
	continuous limit this argument no longer holds and it is therefore far
	easier to produce deviations from the modified FT.

\begin{figure}
		\centering
		\begin{minipage}{0.48\textwidth}
			\centering
			\includegraphics[width=\textwidth]{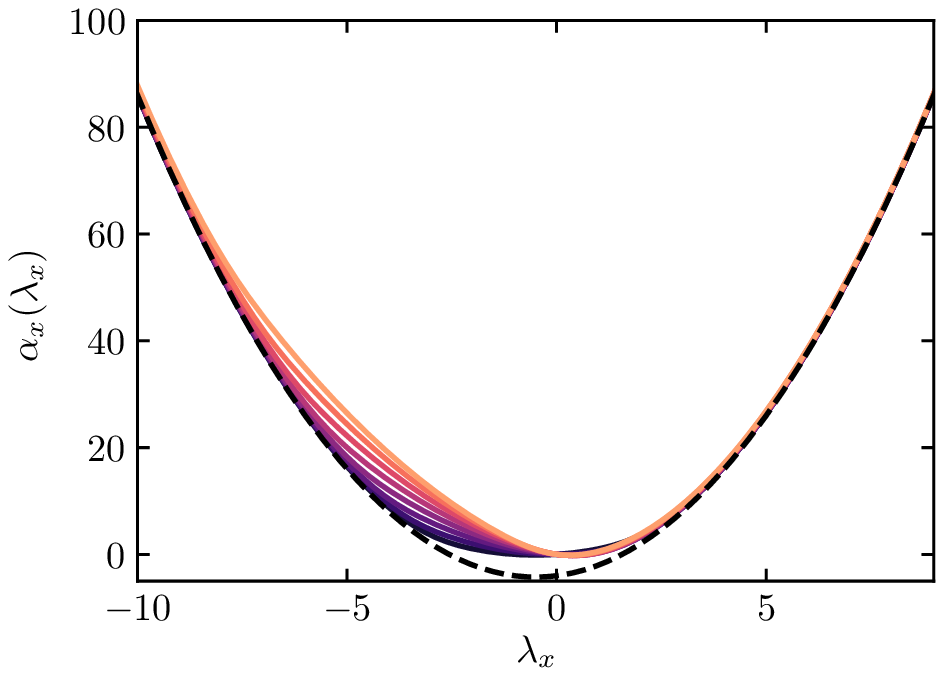}
		\end{minipage}
		\begin{minipage}{0.48\textwidth}
			\centering
			\includegraphics[width=\textwidth]{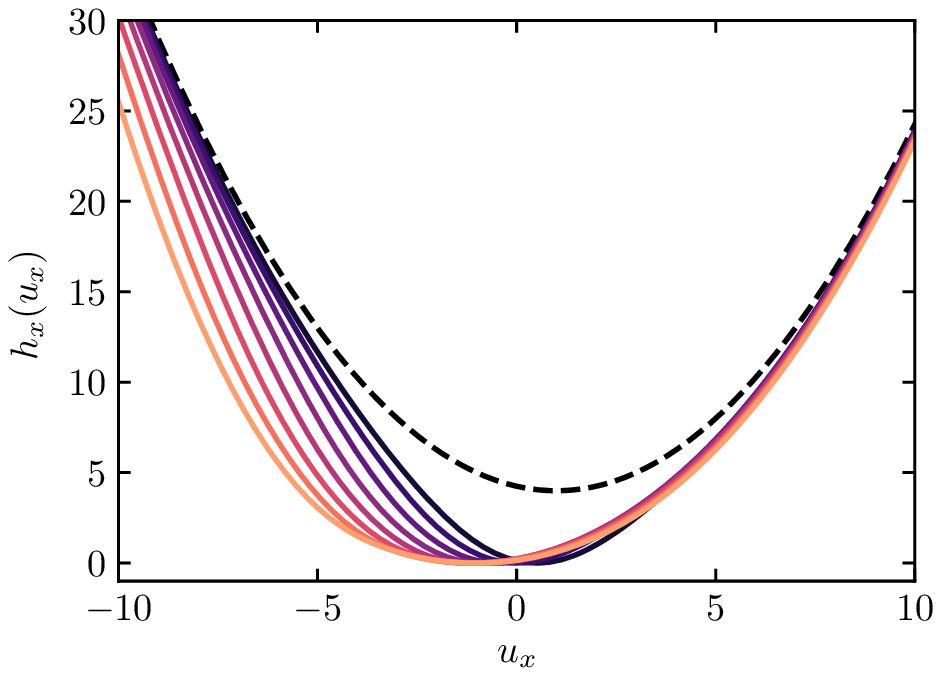}
		\end{minipage}
		\caption{Generating function $\alpha_{x}(\lambda_{x})$ and rate
			function $h_{x}(u_{x})$
			for the two ring system with coupling potential $V=4
			\cos(x_{1} +x_{2})$. The force on the visible degree of freedom
			is fixed at $f_{1}=1$, the force on the hidden degree of freedom varies from
		$f_{2}=0$ (dark) to $f_{2}=7$ (light).  }
		\label{fig:cont_genfunc}
	\end{figure}

	We illustrate this kind of behavior by using the potential energy $V(\xv) =
	V_{0} \cos(x_{1} + x_{2})$. The resulting generating functions and
	rate function are shown in \fref{fig:cont_genfunc}. The symmetric bounds on
	both functions are plotted as dashed lines. Interestingly, for this choice
	of interaction potential they become tight for the extreme values. 

	The comparison of the rate functions for different coupling strengths reveals
	that stronger interaction between the two degrees of freedom leads to a
	decrease of the mean traveled distance. If the interaction is strong enough
	the visible degree moves against the direction of its driving force. The
	asymptotic behavior, however, is not affected by the coupling.

\begin{figure}
		\centering
		\includegraphics[width=0.48\textwidth]{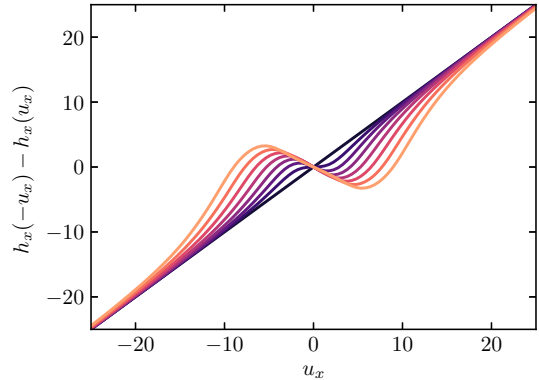}
		\caption{Antisymmetric part of the rate function for the parameter
			values shown in \fref{fig:cont_genfunc}. If the hidden degree of
			freedom is not driven, the exact FT holds for the
			traveled distance of the observable colloid, since this quantity is in
			this case proportional to the total entropy production. If the
			hidden degree of freedom is driven, the FT holds
		only  asymptotically for $|u_{x}| \rightarrow \infty$.}
			\label{fig:cont_fluk}
	\end{figure}

	For the antisymmetric part of the rate function shown
	in~\fref{fig:cont_fluk}, the interaction even leads to negative slopes at the origin if the
	visible degree of freedom moves against its driving force, while the
	asymptotic slope reaches the same value independently of the coupling
	strength since it only depends on the driving of the visible degree of freedom.

	\subsection{Reassessing the interpretation of the experimental data
	from~\cite{mehl_role_2012}}
	\label{sub:reassessing}
	
	So far, our analysis of the rate function has shown that a modified FT holds
	asymptotically for large observation times and large apparent
	entropy production. The experiments and simulations
	in~\cite{mehl_role_2012}, however, hinted at the validity of such a FT for
	finite times. The aim of this section is to put these findings in
	perspective.

	First of all, one should keep in mind that the nature of the FT poses an
	inherent difficulty to experiments and, to lesser extent, also to
	simulations, since it states that events of negative apparent entropy
	production are exponentially suppressed. Therefore one needs to sample
	exponentially many events in order to expand the range in which the FT is
	tested. This range is typically limited to single digit multiples of
	$\kb$. The parameters used for the experiments in~\cite{mehl_role_2012}
	were such that one rotation of the observable colloid produces several
	hundred $\kb$ of apparent entropy; but since the antisymmetric part of the
	distribution can only be evaluated in a much smaller range, this means that
	the events entering in the modified FT reported in~\cite{mehl_role_2012}
	are essentially only those for which the colloid is only moving a few degrees along the
	circumference of the ring in positive or negative direction.

	Large deviation theory on the other hand describes the probability
	distribution in the limit of long observation times. As pointed out
	in~\cite{pietzonka_fine-structured_2014}, where a comparable system with
	hidden degrees of freedom has been analyzed, one can refine the results obtained
	from the rate function by also taking into account sub-exponential
	contributions to the distribution of an integrated current.
	These contributions introduce a fine-structured modulation of the
	probability distribution. The periodicity of the fine structure corresponds in
	the case of the two ring system to the amount of produced apparent entropy
	in one rotation.

	For this reason, the experiments in~\cite{mehl_role_2012} could only capture
	the fine-structure and other subdominant transient contributions to the distribution.
	Likewise the simulation results for cases of broken FT presented therein are
	reminiscent of the impact of the fine-structure on the antisymmetric part
	of the distribution presented in~\cite{pietzonka_fine-structured_2014}.
	Therefore, we can assume that the experiments only capture the leading
	order in $\tstot$ of the asymmetry of the probability distribution, which is in fact linear.
	The time-dependence of the slope of the modified FT in~\cite{mehl_role_2012} can be
	explained by the transient contributions to the distribution.

	\section{Conclusion}
	\label{sec:conclusion}

	We have generalized the notion of apparent entropy production introduced
	in~\cite{mehl_role_2012} to arbitrary discrete networks and have shown that
	the extremes of the probability distribution of the apparent entropy
	production always obey a modified FT in the limit of long
	observation times. As demonstrated on the example of the bipartite lattice,
	this relation can be exploited to infer otherwise inaccessible properties,
	like in this case the true entropy $\Ac_{1}$ produced in one rotation of
	the visible degree of freedom, if the microscopic structure of the system
	is known.

	The analysis of the bipartite lattice also shows that the modified
	FT is in some cases not restricted to the extremes of the
	distribution but holds approximately throughout the whole range of apparent
	entropy production. Even though this result resembles the experimental
	findings of~\cite{mehl_role_2012}, where a modified FT was
	observed for finite times, they are inherently different, since the
	argument explaining the modified FT for the bipartite grid
	is only applicable to a small number of states. 

	For a continuous state space, the
	modified FT holds only for the extreme fluctuations. As in the case of the
	bipartite lattice it is also possible to infer the true affinity of the
	observable degree of freedom from the slope of the modified FT. This slope, however,
	is not identical to the one observed for short observation times in
	experiments. It remains an open question whether it is possible to also infer
	information on hidden properties from the finite time modified FT. 

	\appendix
	\section{Characteristic polynomial of the tilted operator}
	Our goal is to find a representation of the characteristic polynomial of
	the tilted Markov operator $\Lc(\lambda)$, that shows why the largest
	eigenvalue is approximately symmetric under the replacement $\lambda
	\rightarrow \lt - \lambda$.

	The characteristic polynomial is defined as
			\begin{equation}
				\chi(\lambda,y) \equiv \det(\Lc(\lambda)-y) = \sum_{\pi \in
				S_{N^{2}}}
				\sgn(\pi) \prod_{n}\left[ \Lc_{n,\pi(n)}(\lambda) - y
				\delta_{n,\pi(n)} \right] \,,
				\label{eq:char_pol_def}
			\end{equation}
	where we used the Leibniz formula to express the determinant as a sum over
	all permutations $\pi$ of the states. Each of these permutations can itself
	be expressed as a set of non overlapping cycles $\pi =
	\{\mathcal{C}_{\pi}^{i}\}$. Since the matrix $\Lc(\lambda)$ has only
	non-zero entries for connections that correspond to transitions with
	non-zero transitions rates, all relevant cycles, i.e.  those that lead to
	a non-zero term in the sum, are elements of the cycle space of the physical
	network that is studied. They can therefore be expressed as a superposition
	of the fundamental cycles $\mathcal{C}_{\beta}$ of the network, i.e.
	\begin{equation}
		\mathcal{C}_{\pi}^{i} = \sum_{\beta} c_{\pi}^{i,\beta}
		\mathcal{C}_{\beta} \,.
	\end{equation}
	By identifying the diagonal part of the product in~\eref{eq:char_pol_def}
	as 
	\begin{equation}
		\mathcal{D}_{\pi} (y) \equiv \prod_{n| \pi(n) = n}  \left[r(n)-y
		\right]\,,
	\end{equation}
	the product of all transition rates along the direction of the permutation
	as 
	\begin{equation}
		\Gamma_{\pi}^{+} \equiv  \prod_{n | \pi(n) \neq n}  w(n
		\rightarrow \pi(n))\,
	\end{equation}
	and the sum of all increments of the integrated current along all cycles of
	the permutation as its effective affinity
	\begin{equation}
		\At_{\pi} \equiv \sum_{n| \pi(n) \neq n} d_{n,\pi(n)} =
		\sum_{n | \pi(n) \neq n}  \ln \frac{\wt(n \rightarrow
		\pi(n))}{\wt(\pi(n) \rightarrow n)}  \,,
	\end{equation}
	the characteristic polynomial simplifies to
	\begin{equation}
		\chi(\lambda, y) = \sum_{\pi \in S_{N^{2}}} \sgn(\pi) \mathcal{D}_{\pi}(y)
		\Gamma_{\pi}^{+}   \exp \left(\lambda \At_{\pi}      \right) \,.
	\end{equation}
	Note that the diagonal part is the only quantity that depends on $y$.

	To highlight the origin of the symmetry that is observed for the largest
	eigenvalue, we perform this sum twice and divide the result by two. The
	second sum runs over all inverse permutations $\pi^{-1}$. Obviously the
	diagonal does not change under this replacement and the product of rates
	along the cycles becomes the product of rates against the cycles
	\begin{equation}
		\Gamma^{-}_{\pi} \equiv \Gamma^{+}_{\pi^{-1} } = \prod_{n | \pi(n) \neq
		n} w(\pi(n) \rightarrow n) \,.
	\end{equation}

	This leaves us with
	\begin{eqnarray}
		\chi(\lambda, y) = \sum_{\pi\in S_{N^{2}}} \sgn(\pi) \mathcal{D}_{\pi}(y)
		\frac{1}{2} \left[ \Gp \exp(\At_{\pi}\lambda) + \Gm \exp(-\At_{\pi}
		\lambda )    \right]    \\
		=\sum_{\pi \in S_{N^{2}}} \sgn(\pi) \mathcal{D}_{\pi}(y) \sqrt{\Gp \Gm} \cosh \left(
		\At_{\pi} \lambda + \frac{1}{2} \Ac_{\pi}     \right)\,,
	\end{eqnarray}	
	where we introduced the true affinity of the permutation
	\begin{equation}
		\Ac_{\pi} \equiv \ln \frac{\Gp}{\Gm}\,.
	\end{equation}

	Since all cycles of the permutation can be decomposed into fundamental
	cycles the same is also true for the effective and actual affinity of the
	permutation, because those are additive quantities. This allows us to write
	the argument of the hyperbolic cosine function in terms of the constants
	$c_{\pi}^{i,\beta}$  
	\begin{equation}
		\fl	\chi(\lambda, y) = \sum_{\pi \in S_{N^{2}}} \sgn(\pi) \mathcal{D}_{\pi}(y) \sqrt{
		\Gp \Gm} \cosh \left[ \sum_{\mathcal{C}_{\pi}^{i} \in \pi} \sum_{\beta}
		c_{\pi}^{i,\beta} \left(\At_{\beta} \lambda + \frac{1}{2} \Ac_{\beta}
		\right)    \right]  \,.
		\label{eq:univ_charpol}
	\end{equation}
	
	In this representation the fluctuation theorem for the total entropy production
	can be explained easily, since in this case all effective affinities of the
	fundamental cycles coincide with the actual affinities and the whole
	characteristic polynomial is therefore symmetric around $\lambda = -1/2$.

	Since every term of the sum is for itself a symmetric function in $\lambda$,
	the whole characteristic polynomial can be approximately symmetric even if
	we do not consider the total entropy production, but if the effective
	affinities match the actual ones for sufficiently many fundamental
	cycles.

	In the special case of the bipartite lattice there are only three different
	possible combinations of $\Ac_{\beta}$ and $\At_{\beta}$, so the sum inside
	the hyperbolic cosine function in~\eref{eq:univ_charpol} reduces to
	\begin{equation}
		\fl \chi(\lambda, y) = \sum_{\pi \in S_{N^{2}}} \sgn(\pi)
		f_{\pi}(y)  \cosh
		\left[\mathfrak{n}_{\pi} \left(\tilde{\mathcal{A}}\lambda +
		\mathcal{A}_{1}/2 \right) + \mathfrak{m}_{\pi} \mathcal{A}_{2}/2
		\right] \,, 
	\end{equation}
	with $f_{\pi}(y) \equiv \sqrt{\Gp \Gm} \mathcal{D}_{\pi}(y)$.  
	The winding numbers $\nf$ and $\mf$ are the sums over the coefficients
	$c_{\pi}^{i, \beta}$ belonging to the visible or hidden fundamental
	cycles respectively, i.e.
	\begin{equation}
		\nf = \sum_{\mathcal{C}_{\pi}^{i} \in \pi} \sum_{
		\beta\,\text{visible}} c_{\pi}^{i,\beta} \quad \text{and} \quad 
		\mf = \sum_{\mathcal{C}_{\pi}^{i} \in \pi} \sum_{
		\beta\,\text{hidden}} c_{\pi}^{i,\beta} \,.
	\end{equation}
	\section*{References}
	\bibliographystyle{bibgen.bst}
	\bibliography{literature.bib}
\end{document}